\documentstyle[12pt]{article}

\title{
\begin{flushright}
{\normalsize Yaroslavl State University\\
             Preprint YARU-HE-95/03\\
             hep-ph/9506429} \\[5mm]
\end{flushright}
\huge \bf Electromagnetic Catalysis of a
                 Neutrino Radiative decay\\
                 or\\
            One More Source of Information on the
            Lepton Mixing Angles?}

\vspace{10mm}

\author{\Large \bf A.A.~Gvozdev, N.V.~Mikheev  and L.A.~Vassilevskaya\\
{\small\it Division of Theoretical Physics, Department of Physics,}\\
{\small\it Yaroslavl State University, Sovietskaya 14,}\\
{\small\it 150000 Yaroslavl, Russian Federation}\\
\vspace{20mm}
presented by {\Large \bf L.A.~Vassilevskaya}}

\date{}

\begin{document}

\maketitle

\begin{abstract}

The radiative decay of ultrarelativistic massive neutrino
$\nu_i \rightarrow \nu_j \gamma$ is investigated in electromagnetic f\/ields
in the framework of the Standard Model with lepton mixing.
Estimates of the decay probability and
``decay cross-section'' for accelerator neutrinos of high energies
in the electric
f\/ield of nucleus permit one to discuss the general possibility of carrying
out the neutrino experiment. Such an experiment
could give unique information on mixing angles in the lepton sector of
the Standard Model which would be almost independent of the specif\/ic
neutrino masses.

\end{abstract}

\vglue 5mm

\begin{center}
{\it Talk given at the XXXth Rencontres de Moriond \\
Electroweak Interactions and Unified Theories \\
Les Arcs, Savoie, France, March 11-18, 1995}
\end{center}

\newpage

\large

\section{}

   It is known that neutrino oscillations are the main source of
information on the mixing angles and neutrino masses. However limitations
on the mixing angles thus obtained rigidly correlate with the neutrino's
mass spectrum. In this work we'd like to suggest one more source of
information on the lepton mixing angles -- namely, the radiative decay of
a high-energy neutrino in an external electromagnetic field of a nucleus
which has virtually no correlation with the neutrino mass spectrum.

All attempts to find a possible manifestation of lepton mixing
have given up to now negative results. It is reasonable to
suppose that, because  of  the
insufficiently high precision achieved in experimental studies of
neutrino-involving processes, the neutrino mass spectrum  appears
degenerate  (the  neutrinos  manifest  themselves   as   massless
particles). At the same time, with massive neutrino the absence
of lepton mixing seems unnatural and is virtually incompatible
with attempts to somehow extend the standard model (SM).
On the other hand, lepton mixing similar to quark mixing, in itself,
does not go beyond the framework of  the  standard  model and may lead to some
interesting physical phenomena, such as
1) rare lepton decays with lepton number
violation of type $\mu \rightarrow e \gamma$,
$\mu \rightarrow e \gamma \gamma$, $\nu_i \rightarrow \nu_j \gamma$
$\nu_i \rightarrow \nu_j \gamma \gamma$, 2) neutrino oscillations.

It is known that the massive neutrino's properties are
sensitive  to  the  medium  it propagates through. It will
suffice to mention the well known problem of the  solar  neutrino
and  the  possibility  of  solving  it, namely, the  mechanism  of
resonance  enhancement  of  neutrino  oscillations  in  substance.
Substance  is  usually  considered  as
medium. We note, however, that medium can also be represented
by an  external  electromagnetic  field, which  can  significantly
influence both the properties of the massive neutrino itself
and the process of its decay
and even  induce  novel  lepton transitions  with flavour violation
$\nu_i \leftrightarrow \nu_j$ ($i \neq j$) [1] ,
forbidden in vacuum.

\section{}

In the papers [2-5] we investigated the  massive  neutrino  radiative  decay
$\nu_i \rightarrow \nu_j \gamma$ ($i \neq j$)
in external electromagnetic fields (an uniform
magnetic field, a field of intensive electromagnetic wave,
a crossed field, a Coulomb field of nucleus).  The effect of gigantic
enhancement of the decay probability by the
external field (electromagnetic catalysis) was discovered.
In particular, one of the results, we obtained, is that the probability of
the ultrarelativistic neutrino decay in external field
is practically independent of the mass of the
decaying neutrino, if only the decay channel is open
($m_i^2 > m_j^2$).
Here we present the most interesting, in our opinion, result of
the high-energy neutrino radiative decay in the Coulomb field
of nucleus.

At present the experimentally accessible strengths of electromagnetic
fields are significantly below the critical strength
($F/F_e \ll 1, F = B, \cal E$, $F_e = m^{2}_{e}/e \simeq 4.41 \cdot 10^{13} \,
G$).
Because of this, field-induced effects are especially marked in the
ultrarelativstic case with the dynamic parameter

\begin{equation}
\chi^2 = e^{2}(pFFp)/m^{6}
\end{equation}

\noindent being not small even for a relatively weak field ($F_{\mu\nu}$ is
the external field tensor, $p_{\alpha}$ is the 4-momentum, $m$ is
the mass of the particle). This is due to the fact that in the
relativistic particle rest frame the field may turn out of order
of the critical one or even higher, appearing very close to the
constant crossed field ($\vec {\cal E} \perp \vec B$, ${\cal E}=B$).
Thus, the calculation in the case of constant crossed field
is relativistic limit of the calculation in an arbitrary
weak field and possesses a great extent of generality.
What is why the result we obtained for the amplitude of the
ultrarelativstic neutrino radiative decay in the crossed field
can be applied to consider the high-energy neutrino radiative
decay in the Coulomb field of nucleus.
Here the non-uniformity of the electric field $\vec{\cal E}$ of the nucleus
isn't substantial, because the loop process is ``local'' with
the characteristic loop dimension $\Delta x \le (E_\nu e {\cal E})^{-1/3}$
being significantly less than the nucleus size at neutrino energy $E_\nu
\ge 100 \, GeV$.

There is no need to give the total cumbersome expressions for the
propagator and amplitude of the charged fermion in the crossed field.
Here we present the main part of the amplitude of neutrino
radiative decay $\nu_i(p) \rightarrow \nu_j(p') \gamma(q)$
in a physically more interesting case of the ultrarelativistic neutrinos

\begin{eqnarray}
{\cal M} & \simeq & \frac{e^2 G_F}{\pi^2} \; (\varepsilon^* \tilde F p)
\left[ (1-x) + \frac{m_j^2}{m_i^2} (1+x) \right]^{1/2} \sum_\ell
K_{i \ell} K_{j \ell}^* \, J(\chi_\ell) , \nonumber \\
J(\chi_\ell) & = & i \int\limits_0^1 dt \; z_\ell \int\limits_0^\infty du \;
\exp \left[ - i \, (z_\ell u + u^3 / 3) \right] ,
\label{eq:ACF} \\
z_\ell & = & 4 \left[ (1+x) (1-t^2) \left( 1 - \frac{m_j^2}{m_i^2} \right)
\chi_\ell \right]^{-2/3} , \nonumber
\end{eqnarray}

\noindent where $F_{\mu \nu}$, $\tilde F_{\mu \nu} = \frac{1}{2}
\varepsilon_{\mu \nu \alpha \beta} F_{\alpha \beta}$ are the external
electromagnetic f\/ield tensor and its dual tensor,
$\chi_\ell = \sqrt{e^2 (pFFp)} / m_\ell^3$
is the so called dynamic parameter,
$m_{l}$ is the mass of the virtual charged lepton, $K$ is
the lepton mixing matrix of Kobayashi-Maskawa type.
$x = \cos \theta$, $\theta$~is the angle between the vector $\vec p$
(momentum of the decaying ultrarelativistic neutrino) and $\vec q_0$
(momentum of the photon in the decaying neutrino rest frame).
It is worth noting that the amplitude~(\ref{eq:ACF})
is a sum of three loop contributions ($\ell = e, \mu,
\tau$), each one being characterized by its
``field form-factor''~$J(\chi_\ell)$.

In an electric field the dynamic parameter can be represented as follows:

\begin{equation}
\chi_\ell \simeq \left( \frac{E_\nu}{m_\ell} \right) \,
\left( \frac{e {\cal E}}{m_\ell^2} \right) \, \sin \alpha ,
\end{equation}

\noindent where $\alpha$ is the angle between the vector of the momentum
$\vec p$ of the decaying neutrino and the electric field
strength~$\vec{\cal E}$.
In a general way, cumbersome numerical calculations are required to f\/ind
the probability. In the limit of super-high neutrino energies ($E_\nu
\ge 1 \, TeV$), however, the situation is drastically simplif\/ied, as at
such neutrino energies the conditions $\chi_e \gg \chi_\mu \gg 1$,
$\chi_\tau \ll 1$ are fulf\/illed in the vicinity of the nucleus.

\noindent Therefore, it is suf\/f\/icient for us to know only the asymptotics
of the function $J(\chi)$:

\begin{eqnarray}
J(\chi) & = & 1 + O(\chi^2), \qquad \chi \ll 1, \nonumber \\
J(\chi) & = & O(\chi^{-2/3}), \qquad \chi \gg 1,
\end{eqnarray}

\noindent so that the amplitude~(\ref{eq:ACF}) is dominated by the contribution
due
to the virtual $\tau$-lepton. The decay probability can be
presented in the following form:

\begin{equation}
w \simeq \frac{\alpha}{4 \pi} \; \frac{G_F^2}{\pi^3} \; E_\nu e^2
{\cal E}^2 \sin^2 \alpha \; \left( 1 - \frac{m_j^4}{m_i^4} \right) \;
| K_{i \tau} K_{j \tau}^* |^2 .
\end{equation}

We note that there is no suppression associated with the smallness of
the mass of the decaying neutrino.
The comparison of this expression with the
well-known expression for the probability of the radiative decay $\nu_i
\rightarrow \nu_j \gamma$ of a high-energy neutrino in vacuum
(see, for example, [6])  demonstrates the enormous enhancing
influence of the external field on the radiative decay

\begin{equation}
R \equiv \frac{w}{w_0} \simeq \frac{512}{9} \;
\left( \frac{m_W}{m_\tau} \right)^4 \;
\left( \frac{E_\nu}{m_i} \right)^2 \;
\left( \frac{e {\cal E}}{m_i^2} \right)^2 \; \sin^2 \alpha .
\end{equation}

As an illustration, let us give a numerical estimate of $R$ in the case
of the decay of a neutrino of energy $E_\nu \sim 1 \, TeV$ in the vicinity
of a nucleus with the atomic number $Z \sim 20$:

\begin{equation}
R \simeq 2 \cdot 10^{+61} \, \left( \frac{1 \, eV}{m_i} \right)^6 \;
\left( \frac{E_\nu}{1 \, TeV} \right)^2 .
\end{equation}

The neutrino radiative decay in substance must result in $\gamma$-quantum
of energy $E_\gamma \sim E_\nu$ being observed as the decay product.
In experiment, this process would appear as inelastic scattering of the
neutrino on the nucleus. Using this expression  for the probability and
taking the nucleus as a uniformly charged sphere of radius $r_N$
we obtain the following expression for the "cross-section" of the
decay of a high energy neutrino in the electric field of a nucleus
with the atomic number $Z$:

\begin{eqnarray}
\sigma & \simeq & \frac{4}{5} \, Z^2 \, \left( \frac{\alpha}{\pi} \right)^3 \;
\frac{G_F^2 E_\nu}{r_N} \left( 1 - \frac{m_j^4}{m_i^4} \right) \,
| K_{i \tau} K_{j \tau}^* |^2 \nonumber \\
& \simeq & 10^{-44} \, Z^2 \, \left( \frac{10^{-12}cm}{r_N} \right) \,
\left( \frac{E_\nu}{1 \, TeV} \right) \, | K_{i \tau} K_{j \tau}^* |^2
\quad (cm^2).
\end{eqnarray}
It is worthwhile noting that the ``cross-section''  we have presented
is, of course, numerically small,
but not so small as not to allow discussion concerning the possibility
of such an experiment in the future.
It is interesting to compare it with the
elastic $\nu_{\mu}e$-scattering cross-section:

\begin{displaymath}
\sigma^{el}(\nu_{\mu}e)=G^2_{\bar M}m_e E_{\nu}\frac{Z}{8\pi}\simeq
\left( \frac{Z}{100} \right)\left( \frac{E_{\nu}}{100\,GeV} \right)
10^{-40}\,cm^2~,
\end{displaymath}
$\sim4\cdot10^3$ events of which were observed by Charm II Collab. The ratio
of these cross-sections does not depend on $E_{\nu}$ and is:

\begin{displaymath}
\frac{\sigma_Z(\nu_{\mu}\to\nu_e\gamma)}{\sigma^{el}(\nu_{\mu}e)}=
\frac{1}{70A^{1/3}} \left( \frac{Z}{100} \right)
\left( 1-m^4_{\nu_e}/m^4_{\nu_{\mu}} \right)\,
\sin^2\Theta_{12}\sim10^{-3}\sin^2\Theta_{12},
\end{displaymath}
where it was put $r_N=r_0A^{1/3}$, $r_0\simeq1/q_0$, $q_0\simeq0.15\,GeV$,
$\Theta_{12}$ is the mixing angle of $\nu_1$ and $\nu_2$.
This value of the ratio corresponds to a few events of a hard $\gamma$-ray
emission by  neutrino beams in the conditions of Charm II. However, the
number of these events is proportional to $E_{\nu}$, therefore at larger
neutrino beam energy (e.g. at LHC) and with larger detectors the number of
emitted high energy photons (with $\omega_{\gamma}\sim E_{\nu}$) can be of
order of hundreds of thousands per year.


\newpage

\begin{thebibliography}{5}

\bibitem{1}
   A.V.Borisov, I.M.Ternov and L.A.Vassilevskaya,
   Phys. Lett. B {\bf 273}, 163 (1991).


\bibitem{2}
   A.A.Gvozdev, N.V.Mikheev and L.A.Vassilevskaya,
   Phys. Lett. B {\bf 289}, 103 (1992).

\bibitem{3}
   A.A.Gvozdev, N.V.Mikheev and L.A.Vassilevskaya,
   Phys. Lett. B {\bf 321}, 108 (1994).

\bibitem{4}
   A.A.Gvozdev, N.V.Mikheev and L.A.Vassilevskaya,
   Phys. Lett. B {\bf 323}, 179 (1994).

\bibitem{5}
   A.A.Gvozdev, N.V.Mikheev and L.A.Vassilevskaya,
   Yad. Phys. {\bf 57}, 124 (1994).

\bibitem{6}
   B.W.Lee and R.E.Shrock,
   Phys. Rev. D {\bf 16}, 1444 (1977).

\end{thebibliography}
\end{document}